\documentclass[
aps,
prl,
superscriptaddress, 
bibnotes, 
twocolumn, 
amssymb, 
10pt
]{revtex4-1}
\usepackage{soul,color}
\usepackage{amsfonts,amstext}
\usepackage{graphicx} 
\usepackage{dcolumn} 
\usepackage{bm}
\usepackage[flushleft]{threeparttable}
\usepackage{url}
\usepackage{amsmath}
\usepackage{amssymb}
\usepackage{braket}
\usepackage{multirow}
\usepackage{array}
\usepackage{url}
\usepackage{booktabs}
\usepackage{float}
\usepackage{hyperref}
\usepackage{mathtools}
\usepackage{MnSymbol}  
\usepackage{enumitem}
\usepackage{setspace}
\usepackage[dvipsnames]{xcolor}
\usepackage{bbold}  
\usepackage{kotex}

\hypersetup{colorlinks=true, urlcolor=blue, citecolor=cyan, pdfborder={0 0 0}}

\newcommand{\cyan}[1]{{\color{cyan}{\em {#1}}}} 
\newcommand{\vv}[1]{{\boldsymbol #1}}

\newcommand{\fig}[1]{Fig.\,\ref{#1}}

\newcommand{\figr}[1]{Figure\,\ref{#1}}

\newcommand{\thickhline}{\noalign {\ifnum 0=`}\fi \hrule height 1pt\futurelet \reserved@a \@xhline}
\newcolumntype{"}{@{\hskip\tabcolsep\vrule width 1pt\hskip\tabcolsep}}
\newcolumntype{P}[1]{>{\centering\arraybackslash}p{#1}}
\newcommand{\nn}{\nonumber \\}
\newcommand{\rom}[1]{\uppercase\expandafter{\romannumeral #1\relax}}
\pagecolor{white}

\begin{document}
\title{Stable topological phase transitions without symmetry indications in NaZnSb$_{1-x}$Bi$_x$}

\author{Jaemo Jung}
\affiliation{Department of Physics, Sungkyunkwan University, Suwon 16419, Korea}
\author{Dongwook Kim}
\affiliation{Department of Materials Science and Engineering, University of Utah, Salt Lake City, Utah 84112, USA}
\author{Youngkuk Kim}
\email{youngkuk@skku.edu}
\affiliation{Department of Physics, Sungkyunkwan University, Suwon 16419, Korea}
\date{\today}

\begin{abstract} 
We study topological phase transitions in tetragonal NaZnSb$_{1-x}$Bi$_x$, driven by the chemical composition $x$. Notably, we examine mirror Chern numbers that change without symmetry indicators. First-principles calculations are performed to show that  NaZnSb$_{1-x}$Bi$_x$ experiences consecutive topological phase transitions, diagnosed by the strong $\mathbb Z_{2}$ topological index $\nu_{0}$ and two mirror Chern numbers $\mu_{x}$ and $\mu_{xy}$. As the chemical composition $x$ increases, the topological invariants ($\mu_{x}\mu_{xy}\nu_{0}$) change from $(000)$, $(020)$, $(220)$, to $(111)$ at $x$ = 0.15, 0.20, and 0.53, respectively. A simplified low-energy effective model is developed to examine the mirror Chern number changes, highlighting the central role of spectator Dirac fermions in avoiding symmetry indicators. Our findings suggest that NaZnSb$_{1-x}$Bi$_{x}$ can be an exciting testbed for the exploration of the interplay between the topology and symmetry.
\end{abstract}

\maketitle

\cyan{Introduction.- } Since the discovery of archetypal topological insulators protected by time-reversal symmetry \cite{hasan2010colloquium, qi2011}, a considerable number of topological materials with potential applications have been discovered. According to the current topological materials databases, \cite{bradlyn2017, vergniory2019, Vergniory21p2105.09954} out of the 24825 materials tested, 4321 are identified as topological (crystalline) insulators and 10007 are identified as topological semimetals.  Along with topological materials, diverse topological phases have been discovered, which are enriched by diverse symmetries such as translation \cite{Fu07p106803, Moore07p121306, cheng2016, song2017}, inversion \cite{turner2010, hughes2011, PhysRevB.76.045302, Ahn18p106403, Jeon22pL121101}, mirror \cite{teo2008surface}, rotation \cite{fu2011tmpa, Fang12p266802, Alexandradinata2014, fang2019new}, or glide mirror \cite{Young15p126803, fang2015, Wang16hourglass, Wieder18wallpaper}, and with or without time-reversal symmetry \cite{Li10dynamical, Burkov2011, bonderson2013, Alexandradinata2014,tokura2019, Elcoro21magnetic}.  Topological phases are also classified based on their order \cite{benalcazar2017, schindler2018, Khalaf2018a, Dumitru2019}, fragility \cite{po2018}, delicacy \cite{Nelson2021}. Furthermore, obstructed \cite{cano2018, Xu2021, cano2022} and noncompact \cite{schindler2021} atomic insulators. They are applicable with outstanding results in various apparatuses, such as  electronic \cite{Benjamin2011, Steinberg2011, kong2011, tokura2019}, spintronic \cite{hsieh2008, zhang2009, chen2009, Garate2010, he2019}, and quantum computer devices \cite{freedman2002,Das2006,Nayak2008,checkelsky2012}.

The remarkable developments in topological band theory could be one of the fundamental reasons for the success in finding topological materials and phases  \cite{kane2013, Bansil2016}. Moreover, topological quantum chemistry, or equivalently, the symmetry-based indicator method \cite{kruthoff2017, po2017complete, bradlyn2017}, has enabled efficient and high-throughput searches for topological materials. The symmetry indicator significantly simplifies the problem of identifying topological states for a given set of materials. Combined with the first-principles calculations based on density functional theory (DFT), band representations at high-symmetry momenta can efficiently indicate the nontrivial band topology, which refers to a characteristic of energy bands that cannot be adiabatically continued to those of any atomic insulator without either closing a gap or breaking a symmetry \cite{zak1982, Serbyn2014, kruthoff2017}. 

Symmetry indicators are a powerful scheme, but their limitations are clear. First, they fail for a specific set of topological phases, referred to as fragile topological phases \cite{po2018}, which have been a subject of intense study \cite{Kooi19p115160, Hang10p205126, Song20p031001, Zhi-Da2020, peri2020experimental}. Moreover, the symmetry indicators intrinsically have a one-to-many nature \cite{Song18p3530}. Multiple stable topological phases exist for the same trivial indicator. Thus, the Berry phases and Wilson-loop calculations should be employed to determine the stable topological phase. This one-to-many nature allows for a disjointed distinction between the topological phase transitions with and without symmetry indicators. In this study, we examine a class of topological phase transitions that cannot be found from the symmetry indicators. These symmetry-uncaught topological phase transitions can occur because of the lack of symmetry to discern the topological phase transition in terms of symmetry representation \cite{Zhou18p241104, hsu2019purely}.  However, the detailed process of topological phase transitions to avoid symmetry indication remains unexplored.

In this paper, we present a case study of a stable topological phase transition that occurs without symmetry indications. We perform first-principles calculations to study the topological phase transitions in NaZnSb$_{1-x}$Bi$_{x}$ driven by the chemical composition $x$, diagnosed by two mirror Chern numbers $\mu_{x}$ and $\mu_{xy}$ and the strong $\mathbb{Z}_{2}$ topological index $\nu_{0}$.  $(\mu_{x}\mu_{xy}\nu_{0})$ changes from $(000)$, $(020)$, $(220)$, to $(111)$ at $x$ = 0.15, 0.20, and 0.53, respectively. Among these, the topological phase transitions from (000) to (020) and from (020) to (220) occur within the same (trivial) symmetry indicators, thereby uncaught from the symmetry indicators. We build a simplified effective model to demonstrate a mirror Chern number change between the bands with the same symmetry representation, forbidding symmetry indication. We find that symmetry plays a role in the phase transition by providing a constraint on the positions of Dirac fermions and spectator Dirac fermions \cite{PhysRevLett.61.2015, Hatsugai1996, Watanabe2010} in momentum space.

\begin{figure}[t]
\includegraphics[width=0.48\textwidth]{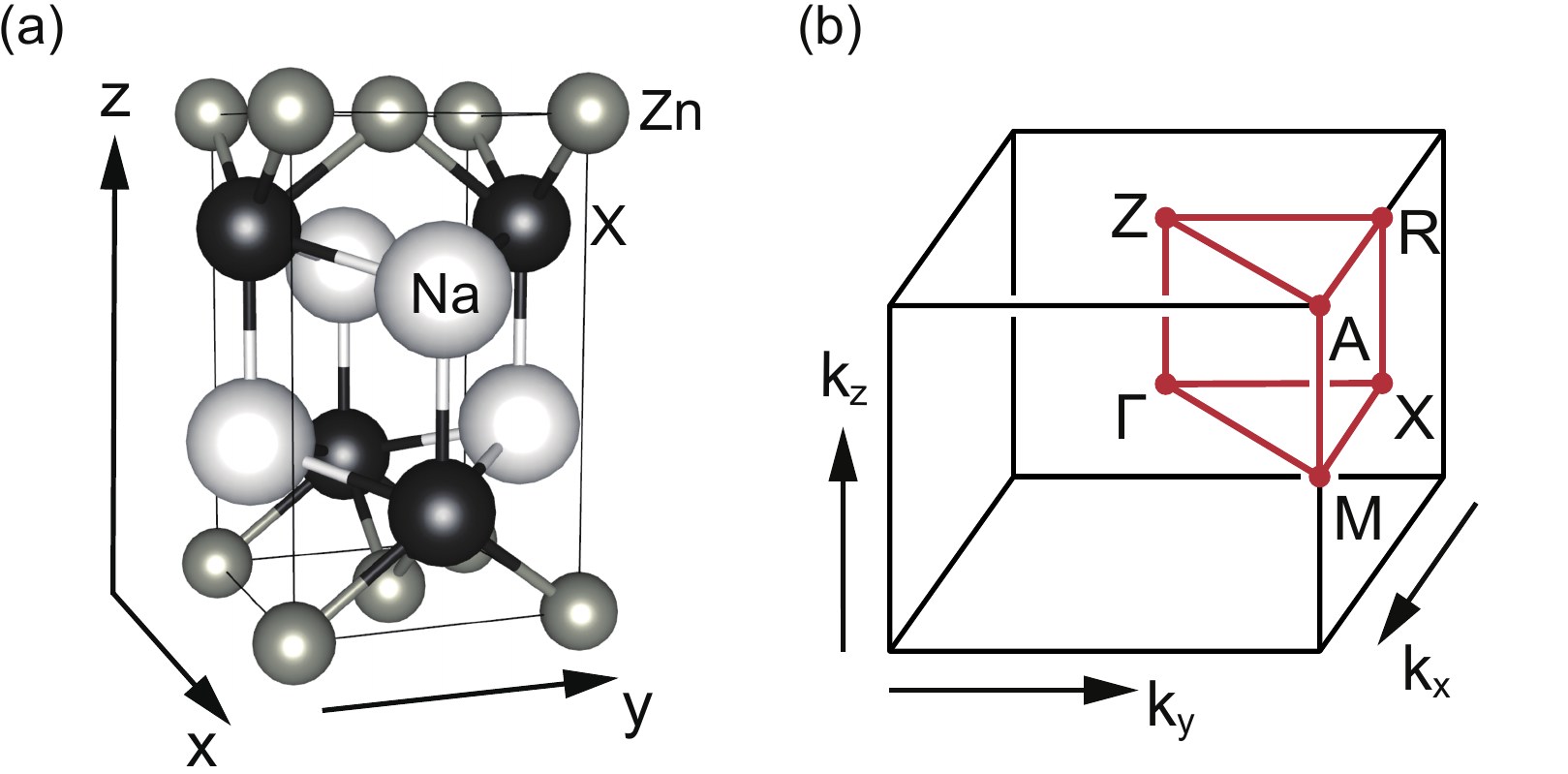} 
\caption{\label{fig:as}
(a) Crystal structure of matlockite-type NaZn$X$ ($X$ = Sb and Bi) in space group $P4/nmm$ (\# 129). Na, Zn, and $X$ atoms are colored by white, grey, and black, respectively. Unit cell is represented by a solid (black) box. (b) Corresponding tetragonal first Brillouin zone. High-symmetry momenta are colored by red.
}
\end{figure}

\cyan{Crystal structure and symmetries.-} \figr{fig:as}(a) shows the crystal structure of NaZn$X$ ($X=$ Bi, Sb) in the space group $P4/nmm$ (\#129). The system comprises Na-$X$ staggered-square sublattices and Zn planar square sublattices, which are placed between the Na-X bilayers. The $P4/nmm$ space group has three generators - two  screw rotations $\{C_{4z}\vert\tfrac{1}{2}\tfrac{1}{2}0\}$ and $\{C_{2x}\vert\tfrac{1}{2}\tfrac{1}{2}0\}$ and spatial inversion $\{\mathcal P\vert\tfrac{1}{2}\tfrac{1}{2}0\}$.  $C_{4z}$ and $C_{2x}$ are a fourfold and twofold rotations about the $\vv z$-axis and $\vv x$-axis, respectively [Fig.\,\ref{fig:as}(a).], and $\{\, g \, \vert\tfrac{1}{2}\tfrac{1}{2}0\}$ ($g = C_{4z}, C_{2x}$, or $\mathcal P$) is a symmetry operator $g$ followed by a fractional translation by half of the primitive unit vectors along the $\vv x$- and $\vv y$-directions. Importantly, there exists $x$-mirror $M_{x}$ and $xy$-glide $G_{xy} = \{M_{xy}\vert\tfrac{1}{2}\tfrac{1}{2}0\}$, which will be employed to evaluate the mirror Chern numbers $\mu_{x}$ and $\mu_{xy}$, respectively. In addition, the system preserves time-reversal symmetry $\mathcal{T}$, enableing the $\mathbb Z_{2}$ topological insulator phase. The first Brillouin zone and the corresponding high-symmetry momenta are shown in \fig{fig:as}(b). Moreover, NaZnSb is an existing material \cite{Jain2013, Jaiganesh2008, Savelsberg1978, Kahlert1976}, whereas NaZnBi is yet to be synthesized.

\begin{figure*}[] 
\includegraphics[width=0.88\textwidth]{
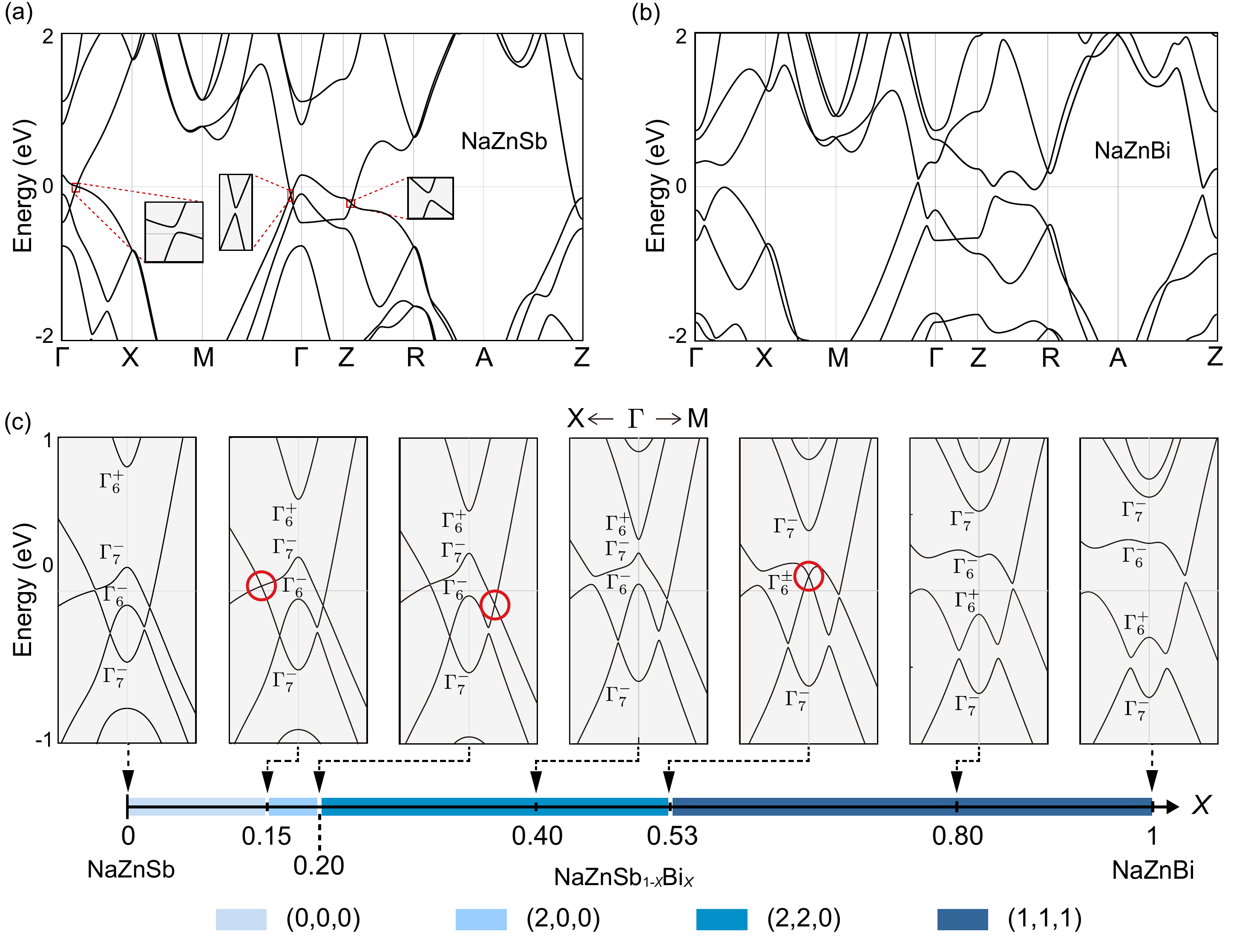}
\caption[]{ 
\label{fig:bs}
(a) DFT bands of NaZnSb ($x$=0). The red rectangles are magnified in the guided grey boxes, showing a direct bandgap. (b) DFT bands of NaZnBi ($x$=1). (c) DFT bands of NaZnSb$_{1-x}$Bi$_x$ for various $x$ and 
topological phase diagram in $x$-space. The bands are drawn near the $\Gamma$ point along with the $X$ and $M$ directions, with the corresponding chemical composition $x$ being indicated in the topological phase diagram. The red circles in the bands highlight the massless (zero-gap) Dirac points, which appear at the topological phase boundaries. Domains in a different colors indicate 
distinct topological phases. 
}
\end{figure*}

\cyan{DFT bands.-} \figr{fig:bs} shows the first-principles electronic energy bands of NaZnSb$_{1-x}$Bi$_x$ calculated for various chemical compositions $x$ using virtual crystal approximation \cite{nordheim1931elektronentheorie, Bellaiche00p7877} \footnote{See Supplementary Materials for the detailed methods.}. A close inspection reveals that a direct bandgap exists throughout the BZ for any $x\in[0,1]$ except for the cases where $x=0.15$, $x=0.20$, and $x=0.53$. In these fine-tuned compositions, the bandgap between the conduction and valence bands vanishes such that it can form a fourfold-degenerate band crossing with linear dispersion, which is dubbed by the Dirac point. Specifically, for the case where $x=0.15$ and $x=0.20$, the Dirac point appears on the $\Gamma-X$  and $\Gamma-M$ lines, respectively, contained in the $M_{x}$- ($G_{xy}$-) invariant $k_{x}=0$ ($k_{x}=-k_{y}$) plane. However, for $x=0.53$, the Dirac point appears at the time-reversal invariant $\Gamma$ point and mediates the band inversion between the $\Gamma_6^+$ and $\Gamma_6^-$ states, as shown in \fig{fig:bs}(c). For any $x \in [0,1]$ other than these critical values, the conduction and valence bands are well separated by a direct bandgap, enabling the evaluation of the topological insulating phase from the occupied bands.  

\cyan{Topological phases.- } The Dirac points accompany a topological phase transition. Using the Wilson loop calculations \cite{Alexandradinata14, Alexandradinata14a, Alexandradinata16b, Alexandradinata16c, Alexandradinata16}, we enumerate two mirror Chern numbers $\mu_{x}$ and $\mu_{xy}$ associated with the $M_x$-mirror and $G_{xy}$-glide on the corresponding invariant planes at $k_x = 0$ and $k_x = -k_y$, respectively \footnote{See Supplementary Materials for the detailed calculations of the mirror Chern numbers}. In addition, the three-dimensional strong $\mathbb{Z}_2$ topological invariant $\nu_{0}$ is calculated using the parity eigenvalues of the occupied bands at eight time-reversal invariant momenta \cite{PhysRevB.76.045302}. As summarized in the bottom panel of \fig{fig:bs}(c), we identify the topological phases characterized by ($\mu_{x},\mu_{xy},\nu_{0}$) = (0,0,0) for $0 \le x < 0.15$, (2,0,0) for $0.15 < x < 0.20$, (2,2,0) for $0.20 < x < 0.53$, and (1,1,1) for $ 0.53 < x \le 1$. Correspondingly, topological phase transitions at $x=0.15$, $x=0.20$, and $x=0.53$ occur owing to the changes in the mirror Chern numbers  $\mu_{x}$ and $\mu_{xy}$ and the strong $\mathbb Z_{2}$ topological index, respectively.

\begin{table}[]
Space group \#129 : P4/nmm\\
\centering
\begin{tabular}{lclclclclclclclcl}
\hline
$\mathbb Z_{2,2,2,4}$ &$\mu_{x,0(\pi)}$ & $\mu_{xy}$ & $g_{z}$ & $g_{xy}$ & $c_{2z}$ & $c_{2x\overline{y}}$ & $c_{4z}$ & $s_{2x}$  \\ \hline
0000 & 0(0)&0&0&0&0&0&0&0 \\ \hline
0000 & 2(0)&0&0&0&0&0&1&1 \\ \hline
0000 & 0(0)&2&0&1&0&1&1&0 \\ \hline
0000 & 2(0)&2&0&1&0&1&0&1 \\ \hline
\end{tabular}
\caption[]{\label{table:SIs} Possible topological crystalline phases corresponding to the trivial $3\mathbb{Z}_2\times\mathbb{Z}_4$ symmetry indicators  ($\nu_{1}\nu_{2}\nu_{3}\nu_{4}$) = (0,0,0,0) in space group \#129. $\mu_{i}$ and $g_{i}$ and $c_{i}$, and $s_{i}$ are the mirror and glide and rotation and screw-resolved topological invariants about the $i$-invariant plane and the $i$-axis, respectively ($i=x, xy, \cdots$). We have applied the results of Ref.\cite{Song18p3530}.}
\end{table}

For completeness, we evaluate the other possible topological crystalline phases allowed in NaZnSb$_{1-x}$Bi$_x$. First, the three-dimensional weak topological insulator phases, characterized by the three weak $\mathbb{Z}_2$ indices $(\nu_1\nu_2\nu_3)$, are turned out to be all trivial $(\nu_1\nu_2\nu_3)=(0,0,0)$ for all gapped phase. In addition, the $\mathbb{Z}_{4}$ index associated with $\mathcal{PT}$ symmetry \cite{Khalaf2018b}, denoted by $\nu_{4}$ is calculated as identical to the $\mathbb{Z}_{2}$ index $\nu_{0}$. Thus, $\nu_{4}=0$ for $x < 0.53$ and $\nu_{4}=1$ for $x >0.53$. Finally, the remaining  topological indices are listed in Table\,\ref{table:SIs}.
Despite the variety, the whole topological crystalline insulator phases are unambiguously determined by the weak indices and the $\mathbb{Z}_{4}$ index  $(\nu_{1},\nu_{2}\nu_{3}\nu_{4})$ along with the two mirror Chern numbers $\mu_{x}$ and $\mu_{xy}$ \cite{Song18p3530}. The mirror Chern number $\mu_{z}$ is associated with the glide $g_{z} = \{M_{z}\vert\tfrac{1}{2}\tfrac{1}{2}0\}$ symmetry. 
The $g_{z}$-invariant plane $k_{z}=0$ hosts four Dirac points at the critical composition $x=0.15$ and $x=0.20$, and the mirror Chern number remains trivial, $\mu_{z}=0$ for $x<0.53$, which is consistent with the symmetry constraint dictated by $\nu_{4}=0$ \cite{Song18p3530}.

\begin{figure}[t]
\includegraphics[width=0.5\textwidth]{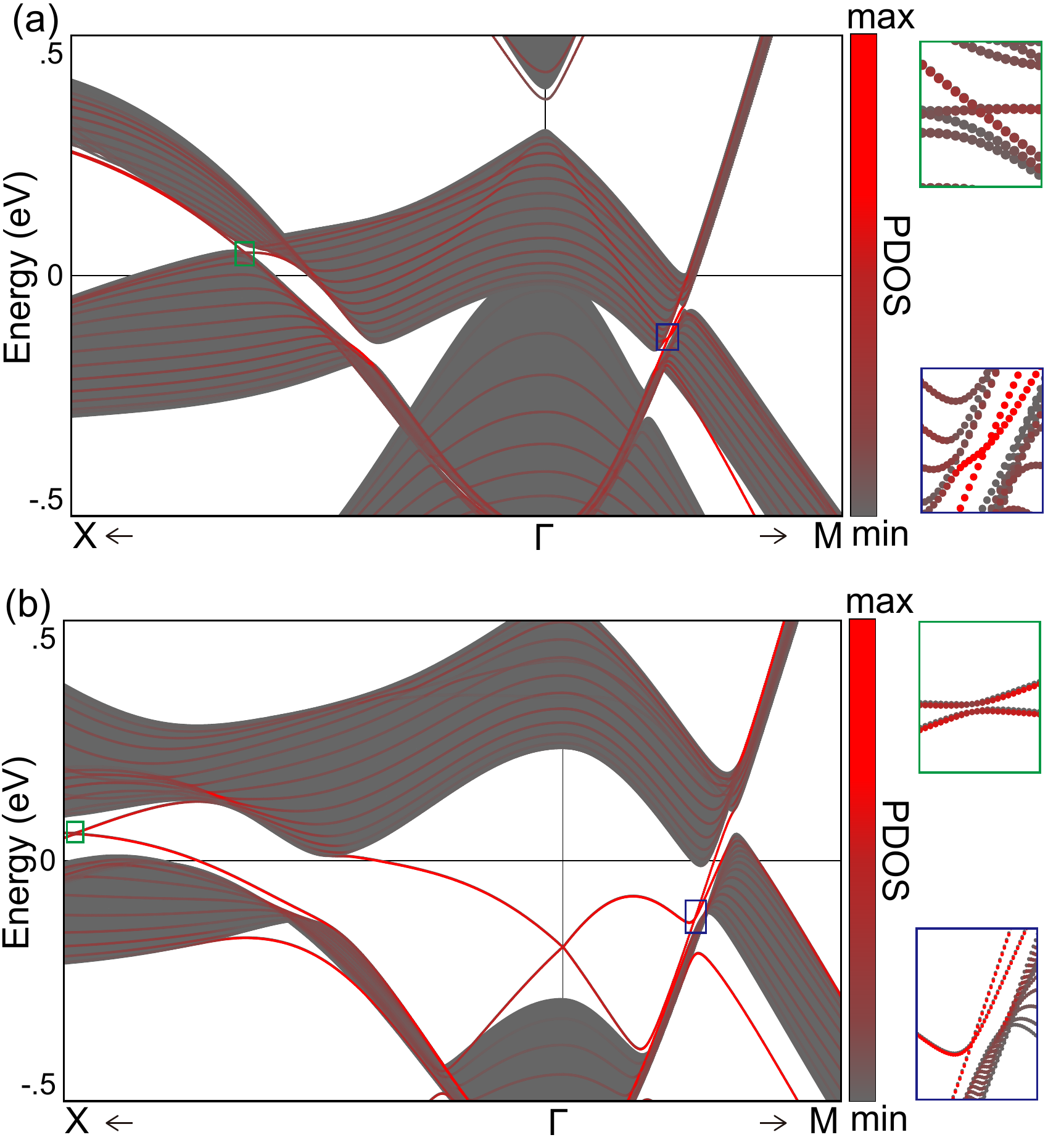}
\caption{\label{fig:ts} 
(001) surface energy spectra of NaZnSb$_{1-x}$Bi$_{x}$ at (a) $x$ = 0.31 and (b) $x$ = 1.00.
Bulk states are denoted in blue, and the surface projection is highlighted in red. 
The magnified view denotes the band crossing (anticrossing) in the surface spectrum for $x=0.31$ ($x$=1.00).
}
\end{figure}

\cyan{Topological surface states.- } The nontrivial topology found in NaZnSb$_{1-x}$Bi$_{x}$ for $x>0.15$ is demonstrated by explicit calculations of  topological surface states.  We prepared a slab geometry of NaZnSb$_{1-x}$Bi$_{x}$ comprising 15 unit cells along the [001]-direction with open boundary conditions imposed on the (001) surface.  
\figr{fig:ts} shows the computed surface states for (a) $x$ = 0.31 and (b) $x$ = 1.00, where $(\mu_{x}\mu_{xy}\nu_{0})=(2,2,0)$ and (1,1,1), respectively. When $x$=0.31, the two surface Dirac points occur because of $\mu_{x}=2$ and $\mu_{xy}=2$ along the high-symmetry $\Gamma-X$ and $\Gamma-M$ lines of the surface BZ, respectively, where the nontrivial mirror planes are projected [\fig{fig:ts}(a)]. For the case of $x=1.00$, on the other hand, the strong topological insulator phase is hosted ($\nu_{0}=1$), leading to the formation of a two-dimensional surface Dirac point occurring at the surface $\Gamma$ point [\fig{fig:ts}(b)]. The calculated surface spectra agree well with the topological phases diagnosed from the bulk topological invariants.

\cyan{Symmetry indicators.-}
After identifying the topological phases of NaZnSb$_{1-x}$Bi$_x$, we evaluate the symmetry indicators. The symmetry indicators proposed in this study is described below. We show that the symmetry indicators proposed in this space group fail to capture the topological phase transitions at $x=0.15$ and $x=20$  by explicitly calculating the symmetry indicators. According to Ref.\,\cite{Song18p3530}, NaZnSb$_{1-x}$Bi$_x$ in space group \#129 contains a set of $3\mathbb{Z}_2\times\mathbb{Z}_4$ symmetry indicators $(\nu_{1}\nu_{2}\nu_{3}\nu_{4})$.  As introduced earlier, the first three indices $\nu_{i=1,2,3}$ are the three-dimensional weak $\mathbb{Z}_2$ topological indices, evaluated from the parity eigenvalues of the occupied bands \cite{PhysRevB.76.045302} and the last index $\nu_4$ is the $\mathcal{P}\mathcal{T}$ symmetric topological invariant, evaluated from $\nu_4 \equiv \sum_{\Gamma_i \in \rm TRIM} \tfrac{n^-_{\Gamma_i} - n^+_{\Gamma_i}}{2}$ (mod 4), where $n^{+(-)}_{\Gamma_i}$ is the number of even- (odd-) parity valence bands at a time-reversal invariant momentum $\Gamma_i$ \cite{Khalaf2018b}. From the first-principles calculations of symmetry representations, we obtain the symmetry indicators $(\nu_{1}\nu_{2}\nu_{3}\nu_{4})=(0000)$ for $x < 0.53$ and  (0001) for $x > 0.53$. Thus, the change in the strong index at $x = 0.53$ is captured by the symmetry indicators, but those at $x=0.15$ and $x=0.20$ are unseen. The absence of a symmetry indication can be attributed to the symmetry representations of the bands. Bacause the topological phase changes via the formation of the Dirac points that reside off the high-symmetry momenta, the symmetry representations of the bands remain the same immediately before and after the Dirac point. Therefore, the failure of the symmetry indicators is inevitable, as evaluated from the symmetry representations.

The failure of symmetry indicators can be comprehended from the symmetry-allowed nature of the Chern numbers. Unlike the symmetry-protected topological phases, the Chern number characterizes a so-called symmetry-forbidden phase, in which symmetries play a role in giving rise to a constraint instead of protection. As shown by Song \textit{et. al.} \cite{Song18p3530}, there are four varieties for a given symmetry indicator in space group \#129 [See Table\,\ref{table:SIs}]. The varieties arise from the two possibilities of the two mirror Chern numbers, that is, $\mu_{i}=0,2$ for $i=x, xy$. which are under the symmetry constraints for the two-fold $C_{2i}$ rotation \cite{Fang12p266802}
\begin{align}
\label{eq:Chern_rot}
e^{i \pi \mu_{i}} = \prod_{n \in \mathrm{occ.}} (-1)^F\prod_{\Gamma_a \in \mathrm{RIM}}\theta_n(\Gamma_a),
\end{align}
where $\theta_n (\Gamma_a) = e^{i (2J_n^{a} +F)\pi/2}$, $J_n^{a}$ is an eigenvalue of the $C_{2i}$ rotation for the $n$-th occupied band at a rotation-invariant momenta (RIM) $\Gamma_{a}$ contained in the mirror-invariant plane, and  $F = 1 (0)$ for a spinful (spinless) system. 
Therefore, the Chern number can be changed by determining $\Delta \mathcal{C}$ from
\begin{align}
\label{eq:chern}
e^{i\pi\Delta\mathcal{C}} = 1,
\end{align}
or equivalently, 
\begin{align} 
\label{eq:chern2}
\Delta\mathcal {C} = 0 ~(\textrm{mod~} 2)
\end{align} 
when the $J_{n}^{a}$ remains the same before and after the variations in Chern number.  Thus, $\mu_{i} = 0$ and $\mu_{i} = 2$ are symmetry-allowed, enabling the varieties of topological phases under the same symmetry structure.

\cyan{Mirror specific four-band model.- } We further resolve the role of symmetry in the change in mirror Chern numbers by constructing an effective Hamiltonian. Let us begin with a generic $4\times4$ Hamiltonian
\begin{align}
\mathcal{H}(\vv k) = \sum_{i,j = x,y,z} h_{ij} (\vv k) \tau_i \sigma_j,
\end{align}
where $\tau_{x,y,z}$ and $\sigma_{x,y,z}$ are the Pauli matrices describing the orbitals and spins, respectively. The $D_{2h}$ point-group symmetries are distilled from the DFT bands responsible for the mirror Chern number change \footnote{See Supple for the derivation of the effective model.}. This leads to the symmetry representations: $\mathcal{T}  = i\sigma_z \mathcal{K}$, $M_{x,y,z} = i\sigma_{x,y,z}$, and $\mathcal P = \mathcal I_{4\times 4}$. Here, $\mathcal{K}$ is the complex conjugation. Under the symmetry constraints 
\begin{equation}
\mathcal{H}(\hat{O}_g \vv k)  = U_g^\dag \mathcal{H}(\vv q ) U_g,
\end{equation}
where $U_g$ and $\hat{O}_g$ are the representation for the symmetry operator $g$ in matrix  and momentum spaces, respectively, 
the effective Hamiltonian on the mirror-invariant plane $k_z = 0$ is obtained as
\begin{align}
\label{eq:H}
\mathcal{H}(\vv k) = A(k_x,k_y) \tau_x + B(k_x,k_y) \tau_z + C(k_x,k_y) \tau_y \sigma_z,
\end{align}
where $A(k_x,k_y) \equiv a_0+a_1 k_x^2+a_2 k_y^2$, $B(k_x,k_y) \equiv b_0+b_1 k_x^2+b_2 k_y^2$, and $C(k_x,k_y) \equiv c_2 k_x k_y$ to the quadratic order in $\vv k = (k_x,k_y)$.
The corresponding energy bands are given by
\begin{align}
E_{\pm} (\vv k) =  \pm\sqrt{A(k_x,k_y)^2 + B(k_x,k_y)^2 + C(k_x,k_y)^2},
\end{align}
for each mirror-sector $\sigma_z = \pm 1$.
The parameters $a_i, b_i$, and $c_2$ ($i$ = 0, 1,and  2) can be 
fine-tuned to critical points, where $A = B = C = 0$. These conditions lead to a  bandgap crossing $E_{+} = E_{-}$ at $\vv k = (\pm\sqrt{-a_0/a_1},0)$ or $\vv k = (0,\pm\sqrt{-a_0/a_2})$. 

\begin{figure}[t]
 \includegraphics[width=0.5\textwidth]{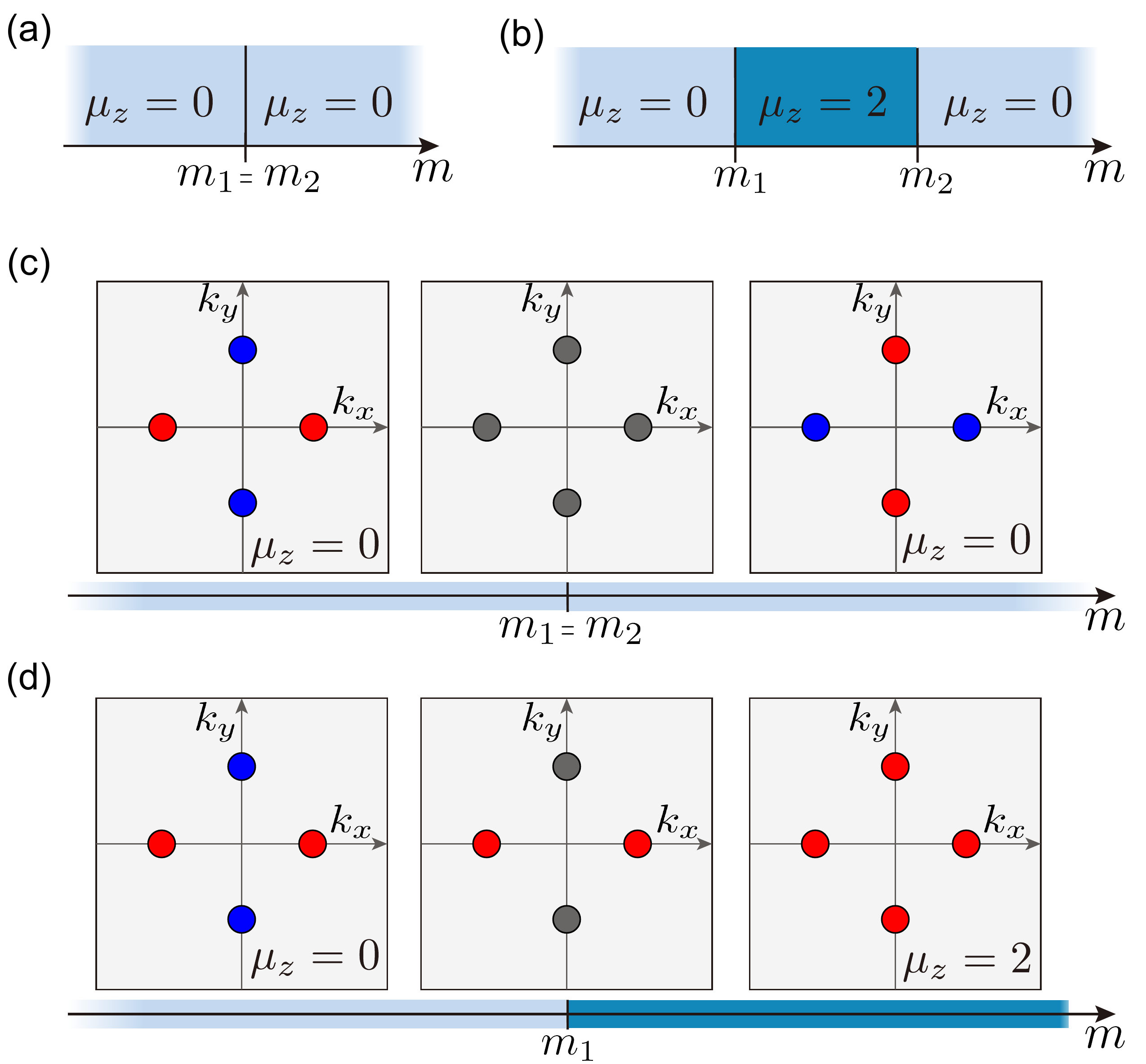}
\caption{
\label{fig:SDC}
Schematic phase diagrams for a topological crystalline phase transition diagnosed by mirror Chern number $\mu_{z}$. (a) Systems with $C_{4z}$. The Dirac and spectator Dirac points that close the band gap simultaneously at $m = m_1 = m_2$ result in the same topological crystalline phases with the mirror Chern number $\mathcal{C}_M = 0$.
(b) Systems without $C_{4z}$.  
The Dirac and spectator Dirac points that close the band gap independently at $m = m_1$ and $m = m_2 \ne m_1$ can mediate a topological phase  transition from $\mu_{z} = 0$ to $\mu_{z} = 2$.
(c-d) Mass inversion of Dirac fermions in the mirror-invariant plane of momentum space: (c) with $C_{4z}$ and (d) without $C_{4z}$. The red and blue circles represent the massive Dirac and the massive spectator Dirac fermions, respectively. The grey circles indicate the massless Dirac fermions. 
}
\end{figure}

The Chern number that characterizes the occupied bands $E_- (\vv k)$ for each mirror-sectors $M_z = \pm i$ is determined by
\begin{align}
\mathcal{C}_{\pm i} = \pm \left\{{\rm sgn}  \left[c_2 \left(\frac{a_0}{a_1} - \frac{b_0}{b_1}\right)\right] - {\rm sgn}\left[c_2\left(\frac{a_0}{a_2} - \frac{b_0}{b_2}\right)\right]\right\},
\end{align}
From which the mirror Chern number $\mu_z$\,\cite{teo2008surface, fu2011tmpa} can be obtained as:
\begin{align}
\label{eq:mcn2}
\mu_{z}
\equiv\,&  \frac{1}{2} (\mathcal{C}_{+i} - \mathcal{C}_{-i}) \nn
=\,& {\rm sgn}\left[c_2 \left(\frac{a_0}{b_0} - \frac{a_2}{b_2}\right)\right] - {\rm sgn}\left[c_2\left(\frac{a_0}{b_0} - \frac{a_1}{b_1}\right)\right].
\end{align}
The nontrivial (trivial) topological crystalline phase indexed by $\mu_{z}=2$ (=0) occurs when $\left(a_0 b_2 - a_2 b_0\right) \left(a_0 b_1 - a_1 b_0\right)<0(>0)$. This equation directly shows that the bandgap crossings define the topological phase transitions between $\mu_z=2$ and $\mu_z=0$.

As illustrated in \fig{fig:SDC}, the results of the effective model provide important insight into the role of symmetries. The $n$-fold rotational symmetry generates $n$ symmetry-related Dirac fermions whose mass is flipped simultaneously during the phase transition. This leads to the variations in Chern number with $n$. We believe that the fraction of $n$ can only be changed when the symmetry is implicitly broken at the representation level, which can be deduced from the symmetry indicators. It is interesting to note the role of the spectator Dirac fermions \cite{PhysRevLett.61.2015, PhysRevLett.61.2015, Hatsugai1996, Watanabe2010}, which refers to the massive Dirac fermions without mass inversion during the transition. Upon restoring a higher-rotational symmetry, such as $C_{4z}$, the topological phase transition becomes trivialized by enforcing the participation of the spectator Dirac fermions. In our case, the  $C_{4z}$-symmetry enforces $a_1 = a_2$ and $b_1 = b_2$, and thus, all the massive Dirac fermions invert the mass simultaneously to nullify the mirror Chern number change. This conforms to the symmetry constraint given by  $C_{4z}$ to the mirror Chern number. It can only change integers that are multiples of four, forbidding two. We believe that this occurs in NaZnSb$_{1-x}$Bi$_{x}$ at $x=0.15$ and $x=0.20$, where four Dirac points occur on the $G_{z}$-invariant $k_{z}=0$ plane without changing the mirror Chern number $\mu_{z}=0$.

\cyan{Conclusions.- }
We have performed a first-principles study on the topological phases of NaZnSb$_{1-x}$Bi$_x$ driven by the chemical composition $x$. We have established the topological phase diagram in $x$-space using symmetry indicators, two mirror  Chern numbers, and the $\mathbb{Z}_{2}$ strong topological index. The phase boundaries are determined to be $x$=0.17, 0.20, and 0.53. 
We focused on analyzing the first two topological phase transitions, which changed the mirror Chern numbers without symmetry indications.
The absence of a symmetry indication is attributed to the intrinsic nature of the Chern numbers and is not based on symmetries. In general, the Chern number can jump by a factor of $n$ without being caught by the $C_{n}$-symmetry, which can be fulfilled by hosting $n$ massless Dirac fermions that mediate the change in the Chern number. 

Our results are scientifically innovative in three aspects. First, the study provides insights into topological phase transitions, uncovering the close interplay between symmetry and topology. Second, we highlight the one-to-many nature of symmetry indicators, suggesting that materials identified as trivial in topology via symmetry inputs can be nontrivial. This may provide opportunities for finding topological materials. Finally, NaZnSb$_{1-x}$Bi$_{x}$ in the tetragonal phase is such an archetypal example that suggests a rich playground for exploring topological phenomena.  

\begin{acknowledgments}
This work was supported by the Korean National Research Foundation (NRF) Basic  Research Laboratory (NRF-2020R1A4A3079707) and 
the NRF Grant number (NRF-2021R1A2C1013871). 
Computational resources were provided by the Korea Institute of Science and Technology Information (KISTI) (KSC-2020-CRE-0108).
\end{acknowledgments}

\end{document}